\documentclass[12pt,a4paper]{article}

\usepackage{a4,graphicx}

\usepackage{amsmath}
\usepackage{amssymb}
\usepackage{amsthm}
\usepackage{bbm}

\usepackage{algorithmic}

\usepackage{maple2e}
 
\DefineParaStyle{Maple Output}
\DefineCharStyle{2D Math}
\DefineCharStyle{2D Output}
\DefineParaStyle{Bullet Item}
\DefineParaStyle{Heading 1}
\DefineParaStyle{Heading 2}
\DefineParaStyle{Warning}
\DefineCharStyle{2D Comment}
\DefineCharStyle{Help Normal}
\DefineCharStyle{Hyperlink}

\newcommand{\Gr}{Gr\"obner }
\newcommand{\cJ}{{\cal J}}

\newcommand{\I}{{\cal I}}

\newcommand{\Z}{\mathbb{Z}}
\newcommand{\K}{{\cal K}}

\newcommand{\M}{{\cal M}}
\newcommand{\R}{{\cal R}}

\newcommand{\lm}{\mathop{\mathrm{lm}}\nolimits}

\newcommand{\lc}{\mathop{\mathrm{lc}}\nolimits}

\newcommand{\Card}{\mathop{\mathrm{Card}}\nolimits}

\newcommand{\Id}{\mathop{\mathrm{Id}}\nolimits}

\newenvironment{algorithm}[1]{
  \begin{center}
    {\bf Algorithm: #1}\\*
     \begin{tabular}{|p{120mm}|} \hline
} {
 \\ \hline
 \end{tabular}
 \end{center}
}

\evensidemargin\oddsidemargin
\advance\textheight-\headheight
\advance\textheight-\headsep
\advance\textheight-\topskip
\title{Computation of Difference \Gr Bases}

\date{}
 
\author{
Vladimir P.Gerdt,\ \ Daniel Robertz
}

\begin{document}

\maketitle

\begin{abstract}
This paper is an updated and extended version of our note \cite{GR'06} (cf.\ also \cite{GR-ACAT}).
To compute difference \Gr bases of ideals generated by linear
polynomials we adopt to difference polynomial rings the involutive algorithm based on Janet-like division.
The algorithm has been implemented in Maple in the form of the package LDA (Linear Difference Algebra) and
we describe the main features of the package. Its applications are
illustrated by generation of finite difference approximations to linear partial differential equations and by reduction of Feynman integrals. We also
present the algorithm for an ideal generated by a finite set of nonlinear difference polynomials. If the algorithm terminates, then it
constructs a \Gr basis of the ideal.
\end{abstract}

%
% ------ Introduction ------
%
\section{Introduction} \label{sec:intro}

Being invented 47 years ago by Buchberger~\cite{Buch65} for algorithmic solving of the
membership problem in the theory of polynomial ideals, the \Gr basis method has become a
powerful universal algorithmic tool for solving various mathematical problems arising in
science and engineering.

Though the overwhelming majority of \Gr basis applications is still found in commutative
polynomial algebra, over the last two decades a substantial progress has also
been achieved in applications
of \Gr bases to noncommutative polynomial algebra, to algebra of differential operators
and to linear partial differential equations (cf., for example, the book~\cite{GBA}). As to the
difference algebra, i.e. algebra of difference polynomials~\cite{Levin'08}, in spite of its conceptual
algorithmic similarity to differential algebra, only a few efforts have been made to
extend the theory of \Gr bases to difference algebra and to exploit their algorithmic
power~\cite{Levin'08,Ch98,Mich'99}.

Recently, three promising applications of difference \Gr bases were revealed:
\begin{itemize}
\item Generation of finite difference approximations to PDEs~\cite{GBM'06,ML'12}.
\item Consistency analysis of such approximations~\cite{GR'10,G'12}.
\item Reduction of multiloop Feynman integrals to the minimal set of basis integrals~\cite{G04}.
\end{itemize}
In this note we describe an algorithm (Section~\ref{sec:algo}) for constructing \Gr bases for linear difference
systems that is an adaptation of the polynomial algorithm~\cite{GB05} to linear difference ideals.
In so doing, we construct a \Gr basis in its Janet-like form (Section~\ref{sec:janetlike}), since this
approach has shown its computational efficiency in the polynomial case~\cite{GB05,InvAlg}. We briefly outline
these efficiency issues in Section~\ref{sec:comp}. The difference form of the algorithm
exploits some basic notions and concepts of difference algebra
(Section~\ref{sec:elements}) as well as the definition of Janet-like \Gr bases
and Janet-like
reductions together with the algorithmic characterization of Janet-like bases
(Section~\ref{sec:janetlike}). Extension of the notion of \Gr basis to nonlinear
difference polynomials, which has not been addressed in \cite{GR'06}, \cite{GR-ACAT},
is briefly described in Section~\ref{sec:nonlinearGB} where we also present the algorithm~\cite{G'12} for construction
of such bases. In Section~\ref{sec:lda} we present our
Maple package LDA for computing \Gr bases
of linear difference ideals, i.e. ideals generated by linear difference polynomials.
The package is a modified version of our earlier package~\cite{Daniel} oriented towards commutative and
linear differential algebra and based on the involutive basis
algorithm~\cite{InvAlg}. The modified version is specialized to
linear difference ideals and uses both Janet and Janet-like
divisions~\cite{GB05} adopted to linear difference
polynomials~\cite{G-ACAT}. In Sections~\ref{sec:diffscheme} and
\ref{sec:Feynman} we illustrate LDA by simple examples of its application to the construction of
finite difference approximations to linear systems of PDEs and to the reduction of Feynman integrals.

%
% ------ Elements of difference algebra ------
%
\section{Elements of difference algebra} \label{sec:elements}

Let $\{y^1,\ldots,y^m\}$ be the set of {\em indeterminates}, e.g., $m$ functions
of $n$ variables $x_1$, \ldots, $x_n$, and $\{ \theta_1,\ldots,\theta_n \}$ be the set of mutually commuting
{\em difference operators (differences)}, i.e.,
$$(\theta_i\circ y^j)(x_1, \ldots, x_n)=y^j(x_1,\ldots,x_i+1,\ldots,x_n).$$
A {\em difference ring $R$ with differences
$\theta_1,\ldots,\theta_n$} is a commutative ring $R$ such that for all
$f,g\in R,\ 1\leq i,j\leq n$,
$$
\begin{array}{l}
\theta_i\theta_j=\theta_j\theta_i,\quad \theta_i \circ (f+g)=\theta_i \circ f+\theta_i \circ g,\\
\theta_i \circ (f\,g)=( \theta_i \circ f) (\theta_i \circ g)\,.
\end{array}
$$
Similarly, one defines a {\em difference field}.

Let $\K$ be a difference field, and $\R:=\K\{y^1,\ldots,y^m\}$ be
the difference ring of polynomials over $\K$ in variables
$$\{\ \theta^\mu \circ y^k\ \mid \mu\in \Z^n_{\geq 0},\,k=1,\ldots,m\ \}\,.$$
Hereafter, we denote by $\R_L$ the set of linear polynomials in $\R$ and use the
notations:
$$
\begin{array}{l}
\Theta=\{\,\theta^\mu\ \mid\ \mu\in \Z^n_{\geq 0}\ \},\quad
\deg_i(\theta^\mu\circ y^k)=\mu_i, \label{theta} \\[0.1cm]
\deg(\theta^\mu \circ
y^k)=|\mu|=\sum_{i=1}^n \mu_i\,.
\end{array}
$$

A {\em difference ideal} is an ideal $\I \subseteq \R$ closed under the action of any
operator from $\Theta$. For $F\subset \R$,
the smallest difference ideal containing $F$
will be denoted by $\Id(F)$. If for an ideal $\I$ there is $F\subset \R_L$ such that $\I=\Id(F)$,
then $I$ is a {\em linear difference ideal}.

A total ordering $\succ$ on the set of
$\theta^\mu \circ y^{\,j}$ is a {\em ranking} if for all $i,j,k,\mu,\nu$ the following
hold:
$$
\begin{array}{l}
\theta_{i} {\theta^\mu \circ y^{\,j}} \succ {\theta^\mu}\circ y^{\,j}\,,\\
\theta^\mu \circ y^{\,j} \succ \theta^\nu \circ y^k \iff
{\theta_i}  {\theta^\mu \circ y^{\,j}}
   \succ {\theta_i}{\theta^\nu} \circ y^k\,.
\end{array}
$$
If $|\mu| > |\nu|$ implies ${\theta^\mu \circ y^{\,j}} \succ {\theta^\nu} \circ y^k$ for all $j$, $k$,
then the ranking is {\em orderly}. If $j > k$ implies ${\theta^\mu} \circ
y^{\,j} \succ {\theta^\nu} \circ y^k$ for all $\mu$, $\nu$, then
the ranking is {\em elimination}.

Given a ranking $\succ$, a linear polynomial $f\in \R_L\setminus \{0\}$
has the {\em leading term}
$a\,\vartheta \circ y^j$, $\vartheta\in \Theta$, $a \in \K$, where $\vartheta \circ y^j$ is maximal
w.r.t.\ $\succ$ among all $\theta^\mu \circ y^k$ which appear with nonzero
coefficient in $f$.
$\lc(f):=a\in \K\setminus \{0\}$ is the {\em leading coefficient}
and $\lm(f):=\vartheta \circ y^{\,j}$
is the {\em leading monomial}.

A ranking acts in $\R_L$ as a {\em monomial order}.
If $F \subseteq \R_L \setminus \{ 0 \}$, $\lm(F)$ will denote the set of the leading monomials and $\lm_j(F)$ will denote
its subset for the indeterminate $y^{\,j}$.
Thus, $$\lm(F)=\bigcup_{j=1}^m \lm_j(F)\,.$$

%
% ------ Janet-like \Gr bases ------
%
\section{Janet-like \Gr bases} \label{sec:janetlike}

Given a nonzero linear difference ideal $\I=\Id(G)$ and a ranking $\succ$,
the ideal generating set
$G=\{g_1,\ldots,g_s\}\subset \R_L$ is a {\em \Gr basis}~\cite{GBA,Mich'99}
of $\I$ if for all $f\in \I\cap \R_L\setminus \{0\}$:
\begin{equation}
\exists \, g\in G, \, \theta \in \Theta\  :\
\lm(f)=\theta \circ \lm(g)\,. \label{GB}
\end{equation}
It follows that $f\in \I\setminus \{0\}$ {\em is reducible modulo $G$}:
$$ f \xrightarrow[g]{} f':=f-\lc(f)\,\theta \circ (g/\lc(g)),\quad f'\in \I\,. $$
If $f'\neq 0$, then it is again reducible modulo $G$, and, by repeating the reduction, in finitely
many steps we obtain
$$\quad f \xrightarrow[G]{}0\,.$$
Similarly, a nonzero polynomial $h\in \R_L$, whose terms are reducible (if any) modulo
a set $F \subset \R_L$,
can be reduced to an irreducible
polynomial $\bar{h}$, which is said to be in {\em normal form modulo $F$}
(denotation: $\bar{h}=NF(h,F)$).

In our algorithmic construction of \Gr bases we shall use a restricted
set of reductions called {\em Janet-like} (cf.~\cite{GB05}) and defined
as follows.

For a finite set $F \subseteq \R_L \setminus \{ 0 \}$ and a ranking $\succ$, we partition
every $\lm_k(F)$ into subsets
labeled by $d_0,\ldots,d_i\in \Z_{\geq 0}$,\ $(0\leq i\leq n)$. Here $[0]_k:=\lm_k(F)$
and for $i>0$ the subset $[d_0,\ldots,d_i]_k$ is defined as
$$
\{u\in \lm_k(F) \mid
d_0=0,d_j=\deg_j(u),1\leq j\leq i \}.
$$
Denote by $h_i(u,\lm_k(F))$ the nonnegative integer
$$\max\{\deg_i(v) \mid u,v\in [d_0, \ldots, d_{i-1}]_k\}-\deg_i(u).$$
If $h_i(u,\lm_k(F))>0$, then $\theta_i^{s_i}$ such that
$$
\begin{array}{l}
s_i:=\min\{\deg_i(v)-\deg_i(u) \mid u,v\in [d_0, \ldots, d_{i-1}]_k,\, \deg_i(v)>\deg_i(u)\}
\end{array}
$$
is called a {\em difference power} for $f\in F$ with $\lm(f)=u$.

\noindent
Let $DP(f,F)$ be the set of difference powers for $f\in F$, and
${\cJ}(f,F):=\Theta \setminus \bar{\Theta}$ be the subset of $\Theta$
with
\[
\bar{\Theta}:=\{\theta^\mu \mid \exists \, \theta^\nu \in DP(f,F)\ :\
\mu-\nu\in \Z^n_{\geq 0} \}.
\]
A \Gr basis $G$ of $I=\Id(G)$ is called {\em Janet-like}~\cite{GB05} if for all
$f \in I\cap \R_L\setminus \{0\}$:
\begin{equation}
\exists \, g\in G, \vartheta
\in {\cJ}(g,G)\  :\ \lm(f)=\vartheta \circ
  \lm(g)\,. \label{JLGB}
\end{equation}
This implies ${\cJ}-$reductions and the ${\cJ}-$normal form $NF_{\cJ}(f,F)$. It is clear
that condition~(\ref{JLGB}) implies condition (\ref{GB}). Note, however, that the converse is generally not
true. Therefore, not every \Gr basis is Janet-like.

The properties of a Janet-like basis are very similar to those of a Janet basis~\cite{InvAlg}, but
the former is generally more compact than the latter. More precisely, let
$GB$ be a reduced
\Gr basis~\cite{GBA}, $JB$ be a minimal Janet basis, and $JLB$ be a minimal
Janet-like basis of the same ideal
for the same ranking. Then we have
\begin{equation}
\Card(GB)\leq \Card(JLB)\leq \Card(JB) \label{cardinalities},
\end{equation}
where $\Card$ abbreviates {\em cardinality}, that is, the number of elements.

Whereas the algorithmic characterization of a \Gr basis is zero redundancy of
all its
$S$-polynomials~\cite{Buch65,GBA}, the algorithmic characterization of a Janet-like basis $G$ is the
following condition~(cf.~\cite{GB05}):
\begin{equation}
\forall g\in G,\ \vartheta\in DP(g,G):\,NF_{\cJ}(\vartheta \circ g,G)=0\,. \label{alg_char}
\end{equation}
This condition is at the root of the algorithmic construction of Janet-like bases as described in the next section.

%
% ------ Algorithm ------
%
\section{Algorithm for Linear Difference Polynomials} \label{sec:algo}

The following algorithm is an adaptation of the polynomial version~\cite{GB05} to linear difference
ideals. It outputs a minimal Janet-like \Gr basis which (if monic, that is, normalized by division
of each polynomial by its leading coefficient) is uniquely defined by the input set $F$ and ranking
$\succ$. Correctness and termination of the algorithm follow from the proof given in~\cite{GB05};
in so doing the displacement of some elements of the intermediate sets $G$ into $Q$ at step~13 provides
minimality of the output basis. The algorithm terminates when the set $Q$ becomes empty in accordance
with~(\ref{alg_char}).

The subalgorithm \textsl{\bfseries{Normal\ Form}}$(p,G,\succ)$ performs the Janet-like reductions (Section~\ref{sec:janetlike}) of the input difference
polynomial $p$ modulo the set $G$ and outputs the Janet-like normal form of
$p$. As long as the intermediate
difference polynomial $h$ has a term Janet-like reducible modulo $G$, the elementary reduction of this term is
done at step~4. As usually in the \Gr bases techniques~\cite{GBA}, the reduction terminates after finitely many
steps due to the properties of the ranking (Section~\ref{sec:elements}).

\begin{algorithm}{\textsl{\bfseries Janet-like \Gr Basis}($F,\succ$)\label{JLB}}
\begin{algorithmic}[1]
\INPUT $F \subseteq \R_{L}\setminus \{0\}$, a finite set;\ $\succ$, a ranking
\OUTPUT $G$, a Janet-like basis of $\Id(F)$
\STATE {\bf choose} $f\in F$ with the lowest $\lm(f)$ w.r.t.\ $\succ$
\STATE $G:=\{f\}$
\STATE $Q:=F\setminus G$
\WHILE{$Q \neq \emptyset$}
  \STATE $h:=0$
  \WHILE{$Q\neq \emptyset$ {\bf and} $h=0$}
    \STATE {\bf choose} $p\in Q$ with the lowest $\lm(p)$ w.r.t.\ $\succ$
    \STATE $Q:=Q\setminus \{p\}$
    \STATE $h:={\bf Normal\ Form}(p,G,\succ)$
  \ENDWHILE
  \IF{$h\neq 0$}
    \FORALL{$g\in G$ such that $\lm(g)=\theta^\mu \circ \lm(h),\ |\mu|>0$}
      \STATE $Q:=Q\cup \{g\}$; \ $G:=G\setminus \{g\}$
    \ENDFOR
    \STATE $G:=G\cup \{ h \}$
    \STATE $Q:=Q\cup \{\,\theta^\beta\circ g \mid g\in G,\
                                    \theta^\beta\in DP(g,G)\,\}$
 \ENDIF
\ENDWHILE
\RETURN $G$
\end{algorithmic}
\end{algorithm}

\begin{algorithm}{\textsl{\bfseries{Normal Form}}$(p,G,\succ)$}
\begin{algorithmic}[1]
\INPUT $p\in \R_{L}\setminus \{0\}$, a polynomial; $G\subset \R_{L}\setminus \{0\}$, a finite set;
       $\succ$, a ranking
\OUTPUT $h=NF_{\cJ}(p,G)$, the ${\cJ}-$normal form of $p$ modulo $G$
\STATE $h:=p$
\WHILE{$h\neq 0$ {\bf and} $h$ has a monomial $u$ with nonzero coefficient $b\in \K$ such that $u$ is ${\cJ}-$reducible modulo $G$}
  \STATE {\bf take} $g\in G$ such that $u=\theta^{\gamma} \circ \lm(g)$ with $\theta^{\gamma}\in {\cJ}(g,G)$
  \STATE $h:=h/b - \theta^{\gamma}\circ (g/\lc(g))$
\ENDWHILE
\RETURN $h$
\end{algorithmic}
\end{algorithm}

An improved version of the above algorithm can easily be derived from
the one for the involutive algorithm~\cite{InvAlg} if one replaces the input involutive division
by a Janet-like monomial division~\cite{GB05} and then translates the algorithm into linear difference algebra.
In particular, the improved version includes Buchberger's criteria adjusted to Janet-like
division and avoids the repeated prolongations $\theta^\beta \circ g$ at step~16 of the algorithm.

%
% ------ Computational aspects ------
%
\section{Computational aspects} \label{sec:comp}

The polynomial version of algorithm \textsl{\bfseries{Janet-like \Gr Basis}} is implemented in its improved form in
C++~\cite{GB05} as a part of the specialized computer algebra system GINV~\cite{ginv}. It has disclosed its high computational efficiency for the standard set of benchmarks\footnote{Cf.\ the web page
{\tt http://invo.jinr.ru}.}. If one compares this algorithm with the involutive one~\cite{InvAlg} specialized
to Janet division, then all the computational merits of the latter algorithm are retained, namely:
\begin{itemize}
\item Automatic avoidance of some useless reductions.
\item Weakened role of the criteria: even without applying any criteria the algorithm is reasonably fast.
By contrast, Buchberger's algorithm without applying the criteria becomes unpractical even for rather
small problems.
\item Smooth growth of intermediate coefficients.
\item Fast search of a polynomial reductor which provides an elementary Janet-like reduction of the given term.
It should be
noted that as well as in the involutive algorithm such a reductor, if it exists, is unique. The fast search is based on the
special data structures called Janet trees~\cite{InvAlg}.
\item Natural and effective parallelism.
\end{itemize}
Though one needs intensive benchmarking for linear difference systems, we have solid
grounds to believe that the above listed computational merits hold also for the difference case.

As this takes place, computation of a Janet-like basis is more efficient than
computation of a Janet basis by the involutive algorithm~\cite{InvAlg}. The
inequality~(\ref{cardinalities}) for monic bases is a consequence of the inclusion~\cite{GB05}:
\begin{equation}
GB\subseteq JLB\subseteq JB\,. \label{subsets}
\end{equation}
There are many systems for which the cardinality of a Janet-like basis is much closer to that of the reduced \Gr basis
than the cardinality of a Janet basis. Certain binomial ideals called toric form an important class of such problems.
Toric ideals
arise in a number of problems of algebraic geometry and closely related to integer programming. For this class
of ideals the cardinality of Janet bases is typically much larger than that of reduced \Gr bases~\cite{GB05}.
For illustrative purposes consider a difference analogue of the simple toric ideal~\cite{GB05,BLR99} generated in the
ring of difference operators by the following set:
$$\{\ \theta_x^7-\theta_y^2\theta_z, \theta_x^4\theta_w-\theta_y^3, \theta_x^3\theta_y-\theta_z\theta_w\ \}\,.$$
The reduced \Gr basis for the degree-reverse-lexicographic ranking with $\theta_x\succ \theta_y\succ
\theta_z\succ \theta_w$ is given by
$$ \{\ \theta_x^7-\theta_y^2\theta_z, \theta_x^4\theta_w-\theta_y^3, \theta_x^3\theta_y-\theta_z\theta_w,
\theta_y^4-\theta_x\theta_z\theta_w^2\ \}\,.$$
The Janet-like basis computed by the above algorithm contains one more element $\theta_x^4\theta_w-\theta_y^3$
whereas the Janet basis adds another six elements to the Janet-like basis~\cite{GB05}.

The presence of extra elements in a Janet basis in comparison with a Janet-like basis is obtained
because of certain additional
algebraic operations. That is why the computation of a Janet-like basis is
more efficient than the computation of a Janet basis.
Both bases, however, contain the reduced \Gr basis as the
internally fixed~\cite{InvAlg} subset of the output
basis\footnote{In the improved
versions of the algorithms.}.
Hence, having any of the bases computed, the reduced \Gr basis is easily extracted without any extra
computational costs.

\section{Nonlinear Difference Polynomials} \label{sec:nonlinearGB}

In this section we follow the paper~\cite{G'12} and define {\em difference standard bases} which generalize the concept of \Gr bases to arbitrary ideals in the ring ${\R}={\K}\{ y^1,\ldots,y^m \}$ of difference polynomials.

A total ordering $\succ$ on the set ${\M}$ of {\em difference monomials}
\[
{\M}:=\{\,(\theta_1\circ y^{1})^{i_1}\cdots (\theta_m\circ y^{m})^{i_m}\mid \theta_j\in \Theta,\ i_j\in \Z_{\geq 0},\ 1\leq j\leq m\,\}
\]
is {\em admissible} if it extends a ranking and satisfies
\[
(\forall\, t\in {\M}\setminus \{1\})\ [t\succ 1]\ \wedge\
(\,\forall\, \theta\in \Theta)\ (\,\forall\, t,v,w\in {\M}\,)\ [\,v\succ w\Longleftrightarrow t\cdot\theta\circ v\succ  t\cdot \theta\circ w\,].
\]

As an example of admissible monomial ordering we indicate the {\em lexicographical monomial ordering}  compatible with a ranking.

Given an admissible ordering $\succ$, every nonzero difference polynomial $f$ has
the {\em leading monomial} $\lm(f)\in {\M}$ with the {\em leading coefficient} $\lc(f)$.
In what follows, every nonzero difference polynomial is to be {\em normalized (i.e., monic)} by
division of the polynomial by its leading coefficient.

If for $v,w\in {\M}$ the equality $w=t\cdot \theta\circ v$ holds with $\theta\in \Theta$ and $t\in {\M}$ we shall say that $v$ {\em divides} $w$ and write $v\mid w$. It is easy to see that this divisibility relation yields a {\em partial order}.

Given a difference ideal ${\I}$ and an admissible monomial ordering $\succ$, a
subset $G\subset {\I}$ is its {\em (difference) standard basis} if $\Id(G)={\I}$ and
\begin{equation}
(\,\forall\, f\in {\I}\,) (\,\exists\, g\in G\,)\ \ [\,\lm(g)\mid \lm(f)\,]\,.
\label{SB}
\end{equation}
As in differential algebra~\cite{Ollivier'90}, if a standard basis is finite it is called \Gr basis.

A polynomial $p\in \R \setminus \{ 0 \}$ is said to be {\em head reducible modulo $q\in \R \setminus \{ 0 \}$ to $r$} if $r=p-m\cdot\theta\circ q$ and $m\in {\M}$, $\theta\in \Theta$ are such that $\lm(p)=m\cdot\theta\circ \lm(q)$. In this case the transformation from
$p$ to $r$ is an {\em elementary reduction} and denoted by ${p}\xrightarrow[q]{} r$.
Given a set $F\subset \R \setminus \{ 0 \}$, $p$ {\em is head reducible modulo $F$} $($denotation: ${p}\xrightarrow[F]{})$  if there is $f\in F$ such that $p$ is head reducible modulo $f$. A polynomial $p$ {\em is head reducible to $r$ modulo $F$} if there is a chain of elementary reductions
\begin{equation}
p\xrightarrow[F]{}p_1\xrightarrow[F]{} p_2\xrightarrow[F]{}\cdots \xrightarrow[F]{}r\,.
\label{red_chain}
\end{equation}
Similarly, one can define {\em tail reduction}. If $r$ in (\ref{red_chain}) and each of its monomials is neither head nor tail reducible modulo $F$, then we shall say that {\em $r$ is in normal form modulo $F$} and write $r=\mathrm{NF}(p,F)$. A polynomial set $F$ with more then one element is {\em interreduced} if
\begin{equation}
(\,\forall f\in F\,)\ [\,f=\mathrm{NF}(f,F\setminus \{f\})\,]\,. \label{interreduce}
\end{equation}

Admissibility of $\succ$, as in commutative algebra,  provides termination of chain (\ref{red_chain}) for any $p$ and $F$. In doing so, $\mathrm{NF}(p,F)$ can be computed by the difference version of a multivariate polynomial division algorithm~\cite{BW'93,CLO'07}. If $G$ is a standard basis of $\Id(G)$, then from the above definitions it follows
\[
 f\in \Id(G) \Longleftrightarrow \mathrm{NF}(f,G)=0\,.
\]
Thus, if an ideal has a finite standard (Gr\"{o}bner) basis, then its construction solves the ideal membership problem as well as in commutative~\cite{BW'93,CLO'07} and differential~\cite{Ollivier'90,Zobnin'05} algebra. The algorithmic characterization of standard bases, and their construction in difference polynomial rings is done in terms of difference $S$-polynomials.

Given an admissible ordering, and monic difference polynomials $p$ and $q$, the polynomial
$$S(p,q):=m_1\cdot \theta_1\circ p-m_2\cdot \theta_2\circ q$$
is called {\em $S$-polynomial} associated to $p$ and $q$ (for $p=q$ we shall say that the {\em $S$-polynomial} is associated with $p)$ if
$$ m_1\cdot \theta_1\circ  \lm(p)=m_2\cdot \theta_2\circ \lm(q)$$
with coprime $m_1\cdot \theta_1$ and $m_2\cdot \theta_2$.

{\em Algorihmic characterization of standard bases:} Given a difference ideal ${\I}\subset \R$ and an admissible ordering $\succ$, a set of polynomials $G\subset {\I}$ is a standard basis of ${\I}$ if and only if $\mathrm{NF}(S(p,q),G)=0$ for
all $S$-polynomials, associated with polynomials in $G$. This result follows from the above definitions in line with the standard proof of the analogous characterization of \Gr bases in commutative algebra~\cite{BW'93,CLO'07} and with the proof of similar characterization of for standard bases in differential algebra~\cite{Ollivier'90}.

Let ${\I}=\Id(F)$ be a difference ideal generated by a finite set $F\subset {\R}$ of difference polynomials. Then for a fixed admissible monomial ordering the following algorithm \textsl{\bfseries{StandardBasis}}, if it terminates, returns a standard basis $G$ of ${\I}$.  The subalgorithm \textsl{\bfseries{Interreduce}} invoked in step~11 performs mutual interreduction of the elements in $\tilde{H}$ and returns a set satisfying (\ref{interreduce}).

Algorithm \textsl{\bfseries{StandardBasis}} is a difference analogue of the simplest version of Buchberger's algorithm (cf.~\cite{Ollivier'90,BW'93,CLO'07}). Its correctness is provided by the above formulated algorithmic characterization of standard bases. The algorithm always terminates when the input polynomials are linear. If this is not the case, the algorithm may not terminate.  This means that the {\bf do while}-loop (steps~2--10) may be infinite as in the differential case~\cite{Ollivier'90,Zobnin'05}. One can improve the algorithm by taking into account Buchberger's criteria to avoid some useless zero reductions in step~5. The difference  criteria are similar to the differential ones~\cite{Ollivier'90}.

\begin{algorithm}{\textsl{\bfseries{StandardBasis}}\,($F,\succ$)\label{StandardBasis}}
\begin{algorithmic}[1]
\INPUT $F\subset {\cal R}\setminus \{0\}$, a finite set of nonzero polynomials;\\ $\succ$, an admissible monomial ordering \\
\OUTPUT $G$, an interreduced standard basis of $\Id(F)$
\STATE $G:=F$
\DOWHILE
  \STATE  $\tilde{H}:=G$
  \FORALL{$S$-polynomials $s$ associated with elements in $\tilde{H}$}
    \STATE $g:=\mathrm{NF}(s,\tilde{H})$
    \IF{$g\neq 0$}
      \STATE $G:=G\cup \{g\}$
    \ENDIF
  \ENDFOR
\ENDDO{$G\neq \tilde{H}$}
\STATE $G:=$\textsl{\bfseries{Interreduce}}\,($G$)
\RETURN $G$
\end{algorithmic}
\end{algorithm}

%
% ------ The Maple Package LDA ------
%
\section{The Maple Package LDA} \label{sec:lda}

The package LDA (abbreviation for {\bf{L}}inear {\bf{D}}ifference
{\bf{A}}lgebra)\footnote{The package LDA is downloadable from the web page {\tt http://wwwb.math.rwth-aachen.de/Janet}}
implements the involutive basis algorithm \cite{InvAlg}
for linear systems
of difference equations using Janet division. In addition, the package
implements a modification of the algorithm
oriented towards Janet-like division~\cite{GB05} and, thus, computes Janet-like
\Gr bases of linear difference ideals. %~\cite{G-ACAT}.

Table~\ref{table:LDA}
collects the most important
commands of LDA. Its main procedure {\tt JanetBasis} converts a
given set of difference polynomials into its Janet basis or
Janet-like \Gr basis form. More precisely, let $\R$ be the
difference ring (cf.\ Section~\ref{sec:elements}) %\cite{G-ACAT}
of polynomials in the variables
$\theta^\mu \circ y^k$, $\mu \in \Z_{\ge 0}^n$, $k = 1$, \ldots, $m$, with
coefficients in a difference field $\K$
containing $\mathbbm{Q}$ for which the field operations can be
carried out constructively in Maple. We denote again by $\R_L$ the set of linear
polynomials in $\R$.
Given a finite generating set $F \subset \R_L$ for a
linear difference ideal $\I$ in $\R$, {\tt JanetBasis}
computes the minimal Janet(-like Gr{\"o}bner) basis $J$ of $\I$
w.r.t.\ a certain monomial order (ranking).
The input for {\tt JanetBasis} consists of the left hand
sides of a linear system of difference equations in the
dependent variables $y^1$, \ldots, $y^m$, e.g., functions of
$x_1$, \ldots, $x_n$. The difference
ring $\R$ is specified by the lists of independent variables
$x_1$, \ldots, $x_n$ and dependent
variables given to {\tt JanetBasis}. The output %of {\tt JanetBasis}
is a list containing the Janet(-like
Gr{\"o}bner) basis $J$ and the lists of independent and dependent
variables.

After $J$ is computed, the involutive/${\cJ}-$normal form of any element of
$\R_L$ modulo $J$ can be computed using {\tt InvReduce}.
Given $p \in \R_L$ representing a residue class $\overline{p}$ of
the difference residue class ring $\R / \I$, {\tt InvReduce}
returns the unique representative $q \in \R_L$ of $\overline{p}$ which
is not involutively/${\cJ}-$reducible modulo $J$.
A $\K$-basis of the vector space $\R_L / (I \cap \R_L)$ is returned by
{\tt ResidueClassBasis} as a list if it is finite or is enumerated by
a formal power series \cite{JanetsApproach}
in case it is infinite. For examples of how to apply
these two commands, cf.\ Section~\ref{sec:Feynman}.

Given an affine (i.e. inhomogeneous) linear system of difference
equations, a call of {\tt CompCond} after the application of
{\tt JanetBasis} returns a generating set of
compatibility conditions for the affine part of the system,
i.e. necessary conditions for the right hand sides of the
inhomogeneous system for solvability.

Moreover, combinatorial devices to compute the
Hilbert series and polynomial and function etc. \cite{Daniel}
are included in LDA.

For the application of LDA to the reduction of Feynman
integrals, a couple of special commands were implemented
to impose further relations on the master integrals:
By means of {\tt AddRelation} an infinite sequence
of master integrals parametrized by indeterminates which
are not contained in the list of independent variables
is set to zero. Subsequent calls of {\tt InvReduce}
and {\tt ResidueClassBasis} take these additional
relations into account (cf.\ Section~\ref{sec:Feynman}).

LDA provides several tools for dealing with
difference operators. Difference operators represented by
polynomials can be applied to (lists of) expressions containing
$y^1$, \ldots, $y^m$
%the dependent variables
as functions of the independent variables.
Conversely, the difference operators can be extracted
from systems of difference equations.
Leading terms of difference equations
%w.r.t.\ chosen rankings
can be selected.

\begin{table*}[htb]
\caption{Main commands of LDA}
\label{table:LDA}
\renewcommand{\arraystretch}{1.2} % enlarge line spacing
\begin{tabular}{@{}ll}
\hline
{\tt JanetBasis}                  & Compute Janet(-like Gr{\"o}bner) basis\\
{\tt InvReduce}                   & Involutive / ${\cJ}-$reduction modulo Janet(-like Gr{\"o}bner) basis\\
{\tt CompCond}                    & Return compatibility conditions for inhomogeneous system\\
{\tt HilbertSeries} etc.          & Combinatorial devices\\
{\tt Pol2Shift}, {\tt Shift2Pol}  & Conversion between shift operators and equations\\
\hline
\multicolumn{2}{@{}l}{Some interpretations of commands for the reduction of Feynman integrals:}\\
\hline
{\tt ResidueClassBasis}           & Enumeration of the master integrals\\
{\tt AddRelation}                 & Definition of additional relations for master integrals\\
{\tt ResidueClassRelations}       & Return the relations defined for master integrals\\
\hline
\end{tabular}
\end{table*}

We consider difference rings containing
shift operators which act in one direction only.
If a linear system of difference equations is given
containing functions shifted in both directions,
then the system needs to be shifted by the
maximal negative shift in order to obtain a
difference system with shifts in one direction only.
However, LDA allows to change the shift direction
globally.

Unnecessary computations of involutive reductions
to zero are avoided using the four involutive criteria
described in \cite{InvAlg,Criteria,UnnecessaryReduction}.
Fine-tuning is possible by selecting the criteria
individually.

The implemented monomial orders/rankings are the
(block) deg\-ree-re\-verse-le\-xi\-co\-gra\-phi\-cal
and the lexicographical one. In the case of more
than one dependent variable, priority of comparison
can be either given to the difference operators
(``term over position'') or to the dependent variables
(``position over term''/elimination ranking).

The ranking is controlled via options given to
each command separately. The other options described above
can be set for the entire LDA session using
the command {\tt LDAOptions} which also allows
to select Janet or Janet-like division.

%
% ------ Generation of finite difference schemes for PDEs ------
%
\section{Generation of finite difference schemes for PDEs} \label{sec:diffscheme}

%% Created by Maple V Release 5.1 (IBM INTEL LINUX)
%% Source Worksheet: Laplace.mws
%% Generated: Thu Aug 11 18:08:46 2005
%\documentclass{article}
%\usepackage{maple2e}
% \def\emptyline{\vspace{12pt}}
%\DefineParaStyle{Maple Output}
%\DefineParaStyle{Maple Output}
%\DefineCharStyle{2D Math}
%\DefineCharStyle{2D Output}
%\begin{document}
%\begin{maplegroup}
%\begin{mapleinput}
%\mapleinline{active}{1d}{restart;}{%
%}
%\end{mapleinput}
%
%\end{maplegroup}

We consider the Laplace equation $u_{xx} + u_{yy} = 0$
and rewrite it as the conservation law
\[
\oint_{\Gamma} -u_y dx + u_x dy = 0.
\]
Adding the integral relations
\begin{eqnarray*}
\int_{x_j}^{x_{j+2}} u_x dx & = & u(x_{j+2}, y) - u(x_j, y),\\
\int_{y_k}^{y_{k+2}} u_y dx & = & u(x, y_{k+2}) - u(x, y_k)
\end{eqnarray*}
and using the midpoint integration method we obtain
the following discrete system:
\begin{equation} \label{eq:LaplaceDiscrete}
\left\{
   \begin{array}{l}
    -(\theta_x - \theta_x \theta_y^2) \! \circ \! u_y  +
(\theta_x^2 \theta_y - \theta_y) \! \circ \! u_x = 0, \\
    2 \triangle h\, \theta_x \! \circ \! u_x   - (\theta_x^2 - 1) \! \circ \! u = 0,   \\
    2 \triangle h\, \theta_y \! \circ \! u_y  - (\theta_y^2 - 1) \! \circ \! u = 0,
   \end{array}
 \right.
\end{equation}
where $\theta_x$ and $\theta_y$ represent the right-shift operators w.r.t.\ $x$ and $y$, e.g.,
$(\theta_x\circ u_y)(x,y)=u_y(x+1,y)$.

We show how to use LDA to find a finite difference scheme
for the Laplace equation:

\medskip

\begin{maplegroup}
\begin{mapleinput}
\mapleinline{active}{1d}{with(LDA):}{%
}
\end{mapleinput}
\end{maplegroup}

\medskip

\noindent
We enter the independent and the dependent variables
for the problem ($ux > uy > u$):

\medskip

\begin{maplegroup}
\begin{mapleinput}
\mapleinline{active}{1d}{ivar:=[x,y]: dvar:=[ux,uy,u]:}{%
}
\end{mapleinput}
\end{maplegroup}

\medskip

\noindent
Next, we translate (\ref{eq:LaplaceDiscrete}) into the input format
of {\tt JanetBasis}. Note that one can in general use
{\tt AppShiftOp} to apply a difference operator given
as a polynomial similar to the ones in (\ref{eq:LaplaceDiscrete})
to a difference polynomial.

\medskip

\begin{maplegroup}
\begin{mapleinput}
\mapleinline{active}{1d}{L:=[2*h*ux(x+1,y)-u(x+2,y)+u(x,y), 2*h*uy(x,y+1)-u(x,y+2)+u(x,y),
2*h*(ux(x+2,y+1)-ux(x,y+1))+2*h*(uy(x+1,y+2)-uy(x+1,y))]:}{%
}
\end{mapleinput}
\end{maplegroup}

\medskip

\noindent
Then we compute the minimal Janet basis of the linear difference ideal
generated by $L$ w.r.t.\ a ranking which compares the
dependent variables prior to the corresponding difference monomials
(``position over term'' order; this ranking is chosen when
using the option $2$ as below).
The least element of this Janet
basis is by construction a difference polynomial which does not
contain any monomial in $ux$ and $uy$ because $ux > uy > u$.

\medskip

\begin{maplegroup}
\begin{mapleinput}
\mapleinline{active}{1d}{JanetBasis(L,ivar,dvar,2)[1][1];}{%
}
\end{mapleinput}

\mapleresult
\begin{maplelatex}
\mapleinline{inert}{2d}{u(x+4,y+2)-4*u(x+2,y+2)+u(x,y+2)+u(x+2,y+4)+u(x+2,y);}{%
\maplemultiline{
 \mathrm{u}(x + 4, \,y + 2) - 4\,\mathrm{u}(x + 2, \,y + 2) + \mathrm{u}(x, \,y + 2) + \mathrm{u}(x + 2, \,y + 4) + \mathrm{u}(x + 2, \,y) }
}
\end{maplelatex}
\end{maplegroup}

\medskip

The computation takes less than one second of time on a
Pentium III (1 GHz).

Dividing this difference polynomial by $4h^2$ we obtain the
following finite difference scheme:
\[
D_j^2(u_{jk})+D_k^2(u_{jk})=0,
\]
where
\[
D_j^2(u_{jk})=\frac{u_{j+2 \, k} - 2 u_{j \, k} + u_{j-2 \, k}}{4
h^2}
\]
and
\[
D_k^2(u_{jk})= \frac{u_{j \, k+2} - 2 u_{j \, k} + u_{j \, k-2}}{4
h^2}
\]
are discrete approximations of the second order partial derivatives
occurring in Laplace's equation.

%\end{document}
%% End of Maple V Output

%%% Local Variables:
%%% mode: latex
%%% TeX-master: "LDA"
%%% End:

%
% ------ Reduction of Feynman integrals ------
%
\section{Reduction of Feynman integrals} \label{sec:Feynman}

%% Created by Maple V Release 5.1 (IBM INTEL LINUX)
%% Source Worksheet: Feynman.mws
%% Generated: Thu Aug 11 18:44:26 2005
%\documentclass{article}
%\usepackage{maple2e}
% \def\emptyline{\vspace{12pt}}
%\DefineParaStyle{Maple Output}
%\DefineParaStyle{Maple Output}
%\DefineCharStyle{2D Math}
%\DefineCharStyle{2D Output}
%\begin{document}
%\begin{maplegroup}
%\begin{mapleinput}
%\mapleinline{active}{1d}{restart;}{%
%}
%\end{mapleinput}
%
%\end{maplegroup}
%\begin{maplegroup}
%\begin{mapleinput}
%\mapleinline{active}{1d}{with(LDA):}{%
%}
%\end{mapleinput}
%\end{maplegroup}
%

In order to demonstrate how to use LDA for the reduction of
Feynman integrals, we consider a simple one-loop propagator type
scalar integral with one massive and another massless particle:
\[
 f(k,n):=I_{k, n} = \frac{1}{i\pi^{d/2}}
 \int  \frac{d^d s}
  {P_{s-q,m}^{k} P_{s,0}^{n}}.
\]
(Here $k$, $n$ are the exponents of the propagators.)

The basis integrals for this example and the corresponding
reduction formulae were found and studied by several authors (see,
e.g.,~\cite{Tarasov,Smirnov}). Here we apply the \Gr basis
method, as implemented in LDA, directly to the
recurrence relations which have the form:
\begin{equation} \label{eq:FeynmanRecurrence}
\left\{ \begin{array}{l}
\, \! \! [d-2k-n - 2m^2 k \mathbf{1}^{+} -\\
\; \! \! n\mathbf{2}^{+}(\mathbf{1}^{-}-q^2+m^2)] \, f(k+1,n+1)=0,\\
\, \! \! [n-k - k \mathbf{1}^{+}(q^2+m^2-\mathbf{2}^{-}) -\\
\; \! \! n\mathbf{2}^{+}(\mathbf{1}^{-}-q^2+m^2)] \, f(k+1,n+1)=0,
\end{array} \right.
\end{equation}
where
\[
\mathbf{1}^{\pm}f(k,n)=f(k\pm 1,n),\
\mathbf{2}^{\pm}f(k,n)=f(k,n\pm 1).
\]
In addition, it is known that
\begin{equation} \label{eq:AdditionalRelation}
f(k+i,n+j)=0 \quad \forall \, i \leq 0 \quad \forall \, j
\end{equation}
which we will take into account later.

\medskip

\begin{maplegroup}
\begin{mapleinput}
\mapleinline{active}{1d}{ivar:=[k,n]: dvar:=[f]:}{%
}
\end{mapleinput}
\end{maplegroup}

\medskip

\noindent
We enter the recurrence relations (\ref{eq:FeynmanRecurrence}):

\medskip

\begin{maplegroup}
\begin{mapleinput}
\mapleinline{active}{1d}{L:=[(d-2*k-n)*f(k+1,n+1)-2*m^2*k*f(k+2,n+1)-n*f(k,n+2)-
n*(m^2-q^2)*f(k+1,n+2), (n-k)*f(k+1,n+1)-k*(q^2+m^2)*f(k+2,n+1)+
k*f(k+2,n)-n*f(k,n+2)-n*(m^2-q^2)*f(k+1,n+2)]:}{%
}
\end{mapleinput}
\end{maplegroup}

\begin{maplegroup}
\begin{mapleinput}
\mapleinline{active}{1d}{JanetBasis(L,ivar,dvar):}{%
}
\end{mapleinput}
\end{maplegroup}

\medskip

\noindent
Again, the computation time is less than one second.
Now, the master integrals are given by:

\medskip

\begin{maplegroup}
\begin{mapleinput}
\mapleinline{active}{1d}{ResidueClassBasis(ivar,dvar);}{%
}
\end{mapleinput}

\mapleresult
\begin{maplelatex}
\mapleinline{inert}{2d}{[f(k,n), f(k,n+1), f(k+1,n), f(k,n+2), f(k+1,n+1), f(k+2,n)];}{%
\maplemultiline{[\mathrm{f}(k, \,n), \,\mathrm{f}(k, \,n + 1), \,
\mathrm{f}(k + 1, \,n), \, \mathrm{f}(k, \,n + 2), \,\mathrm{f}(k + 1, \,n + 1), \,
\mathrm{f}(k + 2, \,n)]}
}
\end{maplelatex}
\end{maplegroup}

\medskip

\noindent
(\ref{eq:AdditionalRelation}) implies additional relations on
the master integrals. (Here, $j$ is recognized as not being
contained in {\tt ivar} and thus serves as a parameter to
define the additional relations.)

\medskip

\begin{maplegroup}
\begin{mapleinput}
\mapleinline{active}{1d}{AddRelation(f(k,n+j)=0,ivar,dvar):}{%
}
\end{mapleinput}
\end{maplegroup}

\medskip

\noindent
The list of master integrals now becomes:

\medskip

\begin{maplegroup}
\begin{mapleinput}
\mapleinline{active}{1d}{ResidueClassBasis(ivar,dvar);}{%
}
\end{mapleinput}

\mapleresult
\begin{maplelatex}
\mapleinline{inert}{2d}{[f(k+1, n), f(k+1,n+1), f(k+2,n)];}{%
\[
[\mathrm{f}(k + 1, \,n), \,\mathrm{f}(k + 1, \,n + 1), \,\mathrm{f}(k + 2, \,n)]
\]
}
\end{maplelatex}
\end{maplegroup}

\medskip

\noindent
Next, we recompute the Janet basis for $m=0$:

\medskip

\begin{maplegroup}
\begin{mapleinput}
\mapleinline{active}{1d}{m:=0: J:=JanetBasis(L,ivar,dvar):}{%
}
\end{mapleinput}
\end{maplegroup}

\medskip

\noindent
For the special case where $m = 0$, we impose the relation
$f(k+i, n) = 0$ for all $i$:

\medskip

\begin{maplegroup}
\begin{mapleinput}
\mapleinline{active}{1d}{AddRelation(f(k+i,n)=0,ivar,dvar):}{%
}
\end{mapleinput}
\end{maplegroup}

\medskip

\noindent
Now, we are left with one master integral:

\medskip

\begin{maplegroup}
\begin{mapleinput}
\mapleinline{active}{1d}{ResidueClassBasis(ivar,dvar);}{%
}
\end{mapleinput}

\mapleresult
\begin{maplelatex}
\mapleinline{inert}{2d}{[f(k+1,n+1)];}{%
\[
[\mathrm{f}(k + 1, \,n + 1)]
\]
}
\end{maplelatex}
\end{maplegroup}

\medskip

\noindent
We reduce $f(k+2,n+3)$ modulo $J$ taking also
the additionally imposed relations on the master
integrals into account. (Here, the option ``F''
lets {\tt InvReduce} return the result in
factorized form.)

\medskip

\begin{maplegroup}
\begin{mapleinput}
\mapleinline{active}{1d}{InvReduce(f(k+2,n+3),J,"F");}{%
}
\end{mapleinput}

\mapleresult
\begin{maplelatex}
\mapleinline{inert}{2d}{-(2*n+4-d+2*k)*(2*n+2-d+2*k)*(2*k+n-d)*(n+3-d+k)*(n+2-d+k)*f(k+1,n+1
)/((n+1)*(2*n-d+4)*n*q^6*k*(d-2*k-2));}{%
\maplemultiline{
 - ((2\,n + 4 - d + 2\,k)\,(2\,n + 2 - d + 2\,k) \, (2\,k + n - d) \, (n + 3 - d + k) \\
(n + 2 - d + k)\,\mathrm{f}(k + 1, \,n + 1))/( (n + 1) \, (2\,n - d + 4)\,n\,q^{6}\,k \,(d - 2\,k - 2)) }
}
\end{maplelatex}
\end{maplegroup}

\medskip

\noindent
Using {\tt ResidueClassRelations} one can display
the relations imposed on the master integrals:

\medskip

\begin{maplegroup}
\begin{mapleinput}
\mapleinline{active}{1d}{ResidueClassRelations(ivar,dvar,[i,j]);}{%
}
\end{mapleinput}

\mapleresult
\begin{maplelatex}
\mapleinline{inert}{2d}{[f(k,n+j), f(k+i,n)];}{%
\[
[\mathrm{f}(k, \, n + j), \,\mathrm{f}(k + i, \,n)]
\]
}
\end{maplelatex}
\end{maplegroup}

\medskip

\noindent
The difference operators occurring in the last result can
be extracted as polynomials in $\delta_k$, $\delta_n$:

\medskip

\begin{maplegroup}
\begin{mapleinput}
\mapleinline{active}{1d}{Shift2Pol(\%,ivar,dvar,[delta[k],delta[n]]);}{%
}
\end{mapleinput}

\mapleresult
\begin{maplelatex}
\mapleinline{inert}{2d}{[delta[n]^j, delta[k]^i];}{%
\[
[{\delta _{n}}^{j}, \,{\delta _{k}}^{i}]
\]
}
\end{maplelatex}
\end{maplegroup}

%\end{document}
%% End of Maple V Output

%%% Local Variables:
%%% mode: latex
%%% TeX-master: "LDA"
%%% End:

%
% ------ Conclusion ------
%
\section{Conclusion} \label{sec:conclusion}

The above presented algorithm \textsl{\bfseries{Janet-like \Gr Basis}} is implemented, in
its improved form, in the Maple package LDA, and can be applied
for generation of finite difference approximations to linear systems of PDEs, to the consistency
analysis of such approximations~\cite{GR'10},
and to reduction of some loop Feynman integrals.

Alternatively, the
\Gr package in Maple in connection with the Ore algebra
package~\cite{  Ch98} can be used to get the same results.

Two of these three applications were illustrated by rather simple
examples. The first difference system (discrete Laplace equation
and integral relations) contains two independent variables $(x,y)$
and three dependent variables $(u,u_x,u_y)$. The second system
(recurrence relations for one-loop Feynman integral) also contains
two independent variables/indices $(k,n)$, but the only dependent
variable $f$. The second system, however, is computationally
slightly harder than the first one because of explicit dependence
of the recurrence relations on the indices and three parameters
$(d, m^2, q^2)$ involved in the dependence on indices.

Dependence on index variables and parameters is an attribute of recurrence
relations for Feynman integrals. Similar dependence may occur in the generation of
difference schemes for PDEs with variable coefficients containing parameters. Theoretically
established exponential and superexponential (depends on the ideal and ordering)
complexity of constructing polynomial \Gr bases implies
that construction of difference \Gr bases is at least exponentially hard in the number of
independent variables (indices). Besides, in the presence of parameters
the volume of computation grows very rapidly as the number of parameters increases.

The reduction of loop Feynman integrals for more than 3 internal
lines with masses is computationally hard for the current version of the package.
One reason for this is that the Maple implementation does not support Janet trees since Maple does not provide
efficient data structures for trees.

Another reason is that in the improved version of the algorithm there is
still some freedom in the selection strategy for elements in $Q$ to be reduced modulo $G$. Though our
algorithms are much less sensitive to the selection strategy than Buchberger's algorithm, the running time
still depends substantially on the selection strategy: mainly because of dependence of the intermediate
coefficients growth on the selection strategy. To find a heuristically good selection strategy
one needs to do intensive benchmarking with difference systems. In turn, this requires an extensive data base
of various benchmarks that, unlike polynomial benchmarks, up to now is missing for difference systems.

\noindent
For the problem of reduction of multiloop Feynman integrals recently some new reduction algorithms have been designed (cf.~\cite{Smirnov'08} and references therein) that exploit special structure of these integrals and by this reason are computationally
much more efficient then the universal \Gr bases method.

The comparison of implementations of polynomial involutive algorithms for Janet bases in Maple and
in C++~\cite{Daniel} shows that the C++ code is of two or three order faster than its Maple
counterpart. Together with efficient parallelization of the algorithm this
gives a real hope
for its practical applicability to problems of current interest in reduction
of loop integrals.

Thus, for successful application of the \Gr basis technique to multiloop
Feynman integrals with masses and to multidimensional PDEs with multiparametric
variable coefficients one has not only to improve our Maple code
but also to implement the algorithms for computing Janet and/or Janet-like difference bases in
C++ as a special module of the GINV software~\cite{ginv} available on the web page {\tt http://invo.jinr.ru}.

As to the algorithm {\textsl{\bfseries{StandardBasis}}, it has not been yet implemented. Another algorithmic
development also aimed at computation of \Gr bases for systems of nonlinear difference polynomials is described in recent paper~\cite{LaScala}.

%
% ------ Achnowledgements ------
%
\section{Achnowledgements}

The first author was supported in part by grants 01-01-00200 and 12-07-00294 from the
Russian Foundation for Basic Research and by grant 3802.2012.2 from the Ministry of Education and Science of the Russian Federation.

\noindent
{\small Vladimir P. Gerdt\\
Laboratory of Information Technologies\\
Joint Institute for Nuclear Research\\
141980 Dubna, Russia\\
Email: {\it gerdt@jinr.ru}
\vskip 0.2cm
\noindent
Daniel Robertz\\
Lehrstuhl B f{\"u}r Mathematik\\
RWTH Aachen\\
52062 Aachen, Germany\\
Email: {\it daniel@momo.math.rwth-aachen.de}
}


\begin{thebibliography}{99}

\bibitem{GR'06} V.~Gerdt, D.~Robertz. {\em Computation of Gr\"{o}bner Bases for Systems of
Linear Difference Equations}. Computeralgebra-Rundbrief Nr.~37, GI\_DMV\_GAMM, 2005, 8--13.

\bibitem{GR-ACAT} V.~P.\ Gerdt, D.\ Robertz.
{\em A Maple Package for Computing \Gr Bases for Linear Recurrence Relations}.
Nucl. Instrum. Methods 559(1), 2006, pp.\ 215--219.
{\tt arXiv:cs.SC/0509070}

\bibitem{Buch65} B.~Buchberger.
{\em An Algorithm for Finding a Basis for the Residue Class Ring
of a Zero-Dimensional Polynomial Ideal}. PhD Thesis,
University of Innsbruck, 1965 (in German).

\bibitem{GBA} B.~Buchberger, F.~Winkler (Eds.)
{\em \Gr Bases and Applications},
Cambridge University Press, 1998.

\bibitem{Levin'08} A.~Levin. {\em Difference Algebra}. Algebra and Applications, vol.~8. Springer, 2008.

\bibitem{Ch98} F. Chyzak.
{\em \Gr Bases, Symbolic Summation and Symbolic Integration}. In book~\cite{GBA},
pp.\ 32--60.

\bibitem{Mich'99} A.~V.\ Mikhalev, A.~B.\ Levin, E.~V.\ Pankratiev, M.~V.\ Kondratieva.
{\em Differential and Difference Dimension Polynomials. Mathematics and Its Applications},
Kluwer, Dordrecht, 1999.

\bibitem{GBM'06} V.~P.\ Gerdt, Yu.~A.\ Blinkov, V.~V.\ Mozzhilkin. {\em \Gr Bases and Generation of
Difference Schemes for Partial Differential Equations}. SIGMA 2, 051, 2006. {\tt arXiv:math.RA/0605334}

\bibitem{ML'12} B. Martin, V. Levandovskyy. {\em Symbolic Approach to Generation and Analysis of Finite Difference
Schemes of Partial Differential Equations}. In: U. Langer, P. Paule (Eds.) Numerical and Symbolic Scientific Computing: Progress and Prospects. Springer, Wien, 2012, pp.\ 123--156.

\bibitem{GR'10} V.~P.\ Gerdt, D.\ Robertz. {\em Consistency of Finite Difference Approximations for Linear PDE Systems and its Algorithmic Verification}. In: S.~M.\ Watt (Ed.) Proceedings of ISSAC 2010, Association for Computing Machinery, 2010, pp.\ 53--59.

\bibitem{G'12} V.~P.\ Gerdt. {\em Consistency Analysis of Finite Difference Approximations to PDE Systems}.  Proceedings of MMCP 2011 (July 3-8, 2011, Star\'{a} Lesn\'{a}, High Tatra Mountains, Slovakia), G. Adam, J. Bu\v{s}a, M. Hnati\v{c} (Eds.), Lect. Notes Comput. Sci. 7175, Springer, Heidelberg, 2012, pp.\ 28--42. {\tt arXiv:math.AP/1107.4269}

\bibitem{G04} V.~P.\ Gerdt.
{\em \Gr Bases in Perturbative Calculations}.
Nucl. Phys. B (Proc. Suppl.) 135, 2004, pp. 232--237. {\tt arXiv:/hep-ph/0501053}

\bibitem{GB05} V. P. Gerdt, Yu. A. Blinkov.
{\em Janet-like Monomial Division}.
Computer Algebra in Scientific Computing / CASC 2005, V. G. Ganzha, E. W.
Mayr, E. V. Vorozhtsov (Eds.), Springer-Verlag, Berlin, 2005, pp. 174--183.
{\em Janet-like \Gr Bases}, ibid., pp.\ 184--195.

\bibitem{InvAlg} V.~P.\ Gerdt.
{\em Involutive Algorithms for Computing \Gr Bases}.
Computational commutative and non-commutative algebraic
geometry. S. Cojocaru, G. Pfister, V. Ufnarovski (Eds.), IOS Press, Amsterdam, 2005, pp.\ 199--225.
{\tt arXiv:math.AC/0501111}

\bibitem{G-ACAT} V.~P.\ Gerdt.
{\em On Computation of \Gr Bases for Linear Difference Systems}.
Nucl. Instrum. Methods 559(1), 2006, pp.\ 211--214. {\tt arXiv:math-ph/0509050}

\bibitem{ginv} V.~P.\ Gerdt, Yu.~A.\ Blinkov. {\em Specialized Computer Algebra System GINV}. Program. Comput. Soft.
34(2), 2008, 112--123.

\bibitem{Daniel} Yu. A. Blinkov, V. P. Gerdt, C. F. Cid, W. Plesken, D. Robertz.
{\em The Maple Package ``Janet'': I. Polynomial Systems}. Computer Algebra in
Scientific Computing / CASC 2003, V. G. Ganzha, E. W. Mayr, and E.
V. Vorozhtsov (Eds.), Institute of Informatics, Technical
University of Munich, Garching, 2003, pp.\ 31--40. Cf.\ also {\tt http://wwwb.math.rwth-aachen.de/Janet}


\bibitem{BLR99} A. M. Bigatti, R. La Scala, L. Robbiano.
{\em Computing Toric Ideals}. J. Symb. Comput. 27, 1999, pp.\ 351--365.

\bibitem{Ollivier'90} F. Ollivier. {\em Standard Bases of Differential Ideals}.
In: S. Sakata (Ed.), Proceedings of  AAECC-8, Lect. Notes Comput. Sci. 508, Springer, New York, 1990, pp.304--321.

\bibitem{BW'93} T. Becker, V. Weispfenning. {\em \Gr Bases: A Computational Approach to
 Commutative Algebra. Graduate Texts in Mathematics}, vol. 141. Springer, New York, 1993.

\bibitem{CLO'07} D.~Cox, J.~Little, D.~O'Shea. {\em Ideals, Varieties and Algorithms. An
 Introduction to Computational Algebraic Geometry and Commutative Algebra}. 3rd edition.
 Springer, New York, 2007.

\bibitem{Zobnin'05} A. Zobnin. {\em Admissible Orderings and Finiteness
    Criteria for Differential Standard Bases}.
In: M. Kauers (ed.) Proceedings of ISSAC'05. Association for Computing Machinery, 2010, pp.\ 365--372.

\bibitem{Criteria} V. P. Gerdt, D. A. Yanovich.
{\em Experimental Analysis of Involutive Criteria}. Algorithmic
Algebra and Logic 2005, A. Dolzmann, A. Seidl, T. Sturm (Eds.), BOD
Norderstedt, Germany, 2005, pp.\ 105--109.

\bibitem{UnnecessaryReduction} J. Apel, R. Hemmecke.
{\em Detecting Unnecessary Reductions in an Involutive Basis Computation}. J. Symb. Comput. 40(4-5), 2005, pp.\ 1131--1149.

\bibitem{JanetsApproach} W. Plesken, D. Robertz.
{\em Janet's approach to presentations and resolutions for
polynomials and linear pdes}.
Archiv der Mathematik 84 (1), 2005, pp.\ 22--37.

\bibitem{Tarasov} O. V. Tarasov.
{\em Reduction of Feynman graph amplitudes to a minimal set of
basic integrals}. Acta Physica Polonica B29, 1998, pp. 2655--2666. {\tt arXiv:/hep-ph/9812250}

\bibitem{Smirnov} V. A. Smirnov.
{\em Evaluating Feynman Integrals},
STMP 211, Springer, Berlin, 2004.

\bibitem{Smirnov'08} A. V. Smirnov.
{\em Algorithm FIRE — Feynman Integral REduction}. JHEP10, 2008, 107. {\tt arXiv:/hep-ph/0807.3243}

\bibitem{LaScala} R. La Scala. \Gr bases and gradings for partial difference ideals.
{\tt arXiv:math.RA/1112.2065}

\end{thebibliography}
\end{document}